\documentclass[%
 reprint,
 amsmath,amssymb,
 aps,
]{revtex4-1}
\usepackage{graphicx,subfigure}
\usepackage{dcolumn}
\usepackage{bm}
\usepackage{hyperref}
\usepackage[mathlines]{lineno}
 
\newcommand{\be}{\begin{equation}}
\newcommand{\ee}{\end{equation}}

\newcommand{\vev}[1]{\langle{#1}\rangle}

\begin{document}

\title{Holography of the Dirac Fluid  in Graphene with two currents}

\author{Yunseok Seo$^1$, Geunho Song$^1$, Philip Kim$^{2,3}$, Subir Sachdev$^{2,4}$ and Sang-Jin Sin$^{1}$}
 \affiliation{$^1$Department of Physics, Hanyang University, Seoul 133-791, Korea.\\
 $^2$Department of Physics, Harvard University, Cambridge, MA 02138, USA. \\
 $^3$Department of Physics and Astronomy, Seoul National University, Seoul 151-747, Korea\\
 $^4$Perimeter Institute for Theoretical Physics, Waterloo, Ontario, Canada N2L 2Y5.}




\date{\today}

\begin{abstract}
Recent experiments have uncovered evidence of the strongly coupled nature of the graphene: 
the Wiedemann-Franz law is violated by up to a factor of 20 near the charge neutral point.
We describe this strongly-coupled plasma by a holographic model in which there are two distinct
conserved U(1) currents. We find that our analytic results for the transport coefficients for  two current model have a significantly improved 
match to the density dependence of the experimental data than the models with only one  current. The additive structure in the  transports coefficients plays an important role. We also suggest the  origin of the two currents. 
\begin{description}
\item[PACS numbers]
 11.25.Tq, 71.10.Hf.
\end{description}
\end{abstract}

\pacs{Valid PACS appear here}
\maketitle


{\bf Introduction:}
It has been argued that graphene near charge neutrality forms a strongly interacting plasma, the Dirac fluid. 
It does not have well-defined quasiparticle excitations, and amenable to a hydrodynamic 
description \cite{MMSS08,FSMS08,MFS08,Foster,MSF09,Mendoza11,Polini14,Vignale15,Polini15,Levitov16}.   
Evidence for such a Dirac fluid has appeared in recent experiments \cite{pkim} on a  
violation of the Wiedemann-Franz  law (WFL) in extremely clean graphene near the charge neutral point: 
the ratio of heat conductivity and electric conductivity,  ${\cal L}=\kappa/T\sigma $, 
was  found to be up to 20 times the Fermi liquid value. 

The simplest hydrodynamic model \cite{HKMS}, with point-like and uncorrelated disorder and a single conserved U(1) current, 
agrees with  the overall experimental trends, but has difficulty capturing the density dependencies of both the 
electrical ($\sigma$) and thermal ($\kappa$) conductivities \cite{Lucas:2015sya}. An alternative hydrodynamic model, the ``puddle'' model, with long-wavelength disorder in the chemical potential and a single conserved U(1) current, led to a better agreement with observations \cite{Lucas:2015sya}, but still left a room for improvement.

In this letter,  we will explore a model with two conserved U(1) currents.  
The idea is that introducing a new neutral current can enhance the transport of the heat  relative to that of the charge. 
 Our model will be formulated in holographic   terms\cite{Maldacena:1997re,Witten:1998qj}, to utilise the recent progress in the developoment of transport calculation in gauge/gravity duality 
 \cite{Blake:2013owa,
 Andrade:2013gsa,
 Donos:2014cya,
 Donos:2014yya,
 Donos:2014oha, Kim:2014bza, Ge:2014aza, Kim:2015sma,
Kim:2015wba, Blake:2015ina, 
Blake:2014yla,
kkss}.
The Dirac fluid in our model is described by an Anti de Sitter (AdS) black hole in 3+1 dimensions, 
the  holographic dual of 2+1 dimensional system at finite temperature. 
 The momentum dissipation is treated  using scalar  fields, 
 which corresponds to weak point-like disorder.  We calculate electric, thermo-electric  power and  thermal conductivities analytically. 
We find that,  under the assumption that the conserved charges $Q_{1}, Q_{2}$ are  proportional to each other,  the theoretical results  for the density dependencies of the electric and heat conductivities can now satisfactorily match the   
the experimental data in the Dirac fluid regime. 

 One possible mechanism for the extra current  is the kinematic constraints  of energy-momentum conservation on the Dirac cone, which reduce the phase space of  electron and hole scattering significantly \cite{Foster}, allowing electrons and holes to form independent currents as far as the relaxation time for mixing between the currents is presumed to be 
is much longer than the Plankian relaxation time $\hbar/k_{B}T$, the time required for hydrodynamic regime at work.  It should be noted, however, that the estimates of electron and hole equilibration times are made in quasiparticle framework \cite{Foster}, whose validity  in hydrodynamic regime is just assumed  here.
We will see that the kinematics on the Dirac cone also provide a reason why the two charge densities can be proportional.


\vskip .2cm
{\bf DC Transport with two $U(1)$ fields :} 
We start from the action $S= \int d^4 x \sqrt{-g}\cal L$ with two gauge fields $A_{\mu}, B_{\mu}$, 
a dilaton field $\phi$ 
   and the scalar fields $\chi_{1},\chi_{2}$ for momentum dissipation: 
\begin{align}\label{action2}
{\cal L}=& R-\frac{1}{2} \left[ (\partial \phi)^2 +\Phi_1 (\phi) (\partial \chi_1 )^2 +\Phi_2 (\phi) (\partial \chi_2 )^2 \right]  \cr
&-V(\phi) -\frac{Z (\phi)}{4} F^2 -\frac{W (\phi)}{4} G^2  ,
\end{align}
where $F={\rm d}A,~~~G={\rm d} B$ and $F^{2}=F_{\mu\nu}F^{\mu\nu}$ etc. 
We also require positivity of $\Phi_{i}(\phi)$, $Z(\phi)$ and $W(\phi)$.  
The action  (\ref{action2}) yields equations of motion:
\begin{align}
R_{\mu \nu} -\frac{1}{2} g_{\mu \nu} {\cal L} =T_{\mu\nu},  \;
    \nabla_{\mu}  ( \sqrt{-g} \Phi_{i}\nabla^{\mu} \chi_{i} ) =0=
     \partial_{\mu} (\sqrt{-g} Z  F^{\mu \nu} ),  \cr  
 \nabla^2 \phi -\sum_{i=1}^{2}   \frac{\Phi'_i }{2} (\phi)(\partial \chi_i )^2   - V'(\phi) - \frac{ Z'(\phi) } {4} F^2  - \frac{W'(\phi) }{4} G^2   =0,  \cr
T_{\mu\nu}=\frac{1}{2} \partial_{\mu}\phi \partial_{\nu} \phi +\sum_{i=1}^{2} \frac{\Phi_{i}}{2}  \partial_{\mu} \chi_{i}  \partial_{\nu} \chi_{i}  
+\frac{Z }{2}F_{\mu}^{\lambda} F_{\nu \lambda} +\frac{W }{2} G_{\mu}^{\lambda} G_{\nu \lambda}.
\end{align} 
We take the ansatz for   metric and the gauge fields  as 
\begin{align}\label{anz2}
ds^2 &= -U(r) dt^2 +\frac{1}{U(r)}dr^2 + e^{ v(r)}(dx^2 +dy^2) \cr
A &=A(r) dt,~~~~B=B(r) dt.
\end{align}
The gauge field $A$ has  the chemical potential and charge density as its   components of its near boundary expansion,
$
A (r)= \mu_1 - q_1/r + \cdots.
$
At  the horizon at $r=r_0$,   $U$ vanishes and $A, B\to 0$ also for the regularity. 
If we take  the following solution,  
$
\chi_{1} = k x,~\chi_{2} =k y,
$
  it  provides  momentum relaxation. 
 From now on, we set $\Phi_1 =\Phi_2 =\Phi$ for simplicity.
The only non-zero components in the Maxwell equations  are that for the 
$tr$-component of the field strengths whose first integral give conserved charges,
 \begin{align}
 Q_1 &= \sqrt{-g} Z(\phi) F^{tr} = Z(\phi) e^{v} A'(r) \cr
 Q_2 &= \sqrt{-g} W(\phi) G^{tr} =W(\phi) e^{v} B'(r).
 \end{align}
One can see that if $e^{v}\sim r^2 $ in asymptotic region, 
$Q_i$ corresponds to the charge density of the boundary field theory.
 To compute the transport coefficients, we  turn on small fluctuations around the background solution: 
\begin{align}
\delta G_{tx} =& t \delta f_1 (r) +\delta g_{tx}(r) , \quad 
\delta G_{rx} = e^{v(r)} \delta h_{rx} (r), \cr
\delta A_{x} =& t \delta f_2 (r) +\delta a(r),  \quad
\delta B_{x} = t \delta f_3 (r) +\delta b(r) , 
\end{align}
as well as $\delta \chi_i(r)$'s. 
We choose the functions $f_i (r)$ as
\begin{align}\label{fanz}
\delta f_1(r) &= -\zeta U(r), \quad
\delta f_2 (r) =- E_{1} +\zeta A(r) , \cr
\delta f_3 (r) &=-E_{2} +\zeta B(r),
\end{align}
such that the time $t$ does not appear in the equations of motion of the fluctuations.
Here, $E_{1}, E_{2}$ are thmo-electric forces   acting on     $J_{1},J_{2}$ respectively and 
$\zeta=-\nabla T/T$.
 From the $A$ field fluctuation equations, the currents are defined by \cite{Donos:2014cya},
\begin{align}
J_1 &=\sqrt{-g} Z (\phi) F^{xr} ,\quad \quad J_2 =\sqrt{-g} W (\phi)G^{xr} \cr
{\cal Q} &= U(r)^2 \frac{d}{dr}\left(\frac{\delta g_{tx} (r)}{U(r)}\right) -A(r) J_1-B(r) J_2.
\end{align} 
Notice that near the boundary, the heat current becomes 
${\cal Q}=T^{tx}-\mu_1 J_1 - \mu_2 J_2$. 
 Moreover, these currents are conserved along radial direction $r$.  
    Therefore their boundary values are related to that of horizon data,  which can be 
    determined from the regularity at the horizon  \cite{Donos:2014yya}: 
\begin{align}\label{horizonexpansion}
\delta a(r) \sim -\frac{E_{1}}{4\pi T} \ln (r-r_0) , \quad
\delta g_{tx} (r) \sim \delta g_{tx}^{(0)} , \cdots. 
\end{align}
Using above horizon behavior we get the boundary current in terms of the external sources:
\begin{align}\label{current}
J_1 &= \left(Z_0 +\frac{e^{-v_0} Q_1^2}{k^2 \Phi_0} \right) E_{1} + \frac{e^{-v_0} Q_1 Q_2}{k^2 \Phi} E_2 + \frac{4 \pi T Q_1}{k^2 \Phi_0} \zeta,  \cr
J_2 &=\left(W_0 +\frac{e^{-v_0} Q_2^2}{k^2 \Phi_0} \right) E_2 + \frac{e^{-v_0} Q_1 Q_2}{k^2 \Phi} E_{1} + \frac{4 \pi T Q_2}{k^2 \Phi_0} \zeta \cr
{\cal Q}&= \frac{4 \pi T Q_1}{k^2 \Phi_0} E_{1} +\frac{4 \pi T Q_2}{k^2 \Phi_0} E_2 + \frac{(4\pi T)^2 e^{v_0}}{k^2 \Phi_0} \zeta.
\end{align}
 The eq. (\ref{current}) can be written in matrix form, $ {J_{i}} = \Sigma_{ij}{E_{j}}$, with $J_{3}={\cal Q}$ and $E_{3}=\zeta$. 
The transport coefficients can be read off from the eq. (\ref{current}) and the definition 
\begin{align}\label{trans_matrix}
\left(\begin{array}{ccc}     
 \sigma_{1} & {\delta} &\alpha_1 T  \\  
\bar \delta & \sigma_{2}&\alpha_2 T \\  
\bar{\alpha}_1 T & \bar{\alpha}_2 T&\bar{\kappa} T 
\end{array}\right) :=\Sigma .
\end{align} 
Notice that the matrix is real and symmetric,  so  that the  Onsager relations hold:
\begin{align}
{\bar \alpha}_i=\alpha_i, ~~~{\bar \delta}=\delta .
\end{align}
 The heat conductivity $\kappa$ is defined by the response of the temperature gradient to the heat current  in the absence of  other currents: setting $J_1$ and $J_2$ to be zero in (\ref{current}), we can express  $E_{1}$ and $E_2$ in terms of $\zeta$. Substituting these to the first line of  (\ref{current}), we get
\begin{align}\label{thermal_cond}
\kappa &= \bar{\kappa} -     \frac{T\bar{\alpha}_1( \alpha_1 \sigma_2 - \alpha_2 {\delta})}{ \sigma_{1} \sigma_2 -\delta \bar{\delta}} -  \frac{T\bar{\alpha}_2(\alpha_2 \sigma_{1} - \alpha_1 \bar{\delta})}{  \sigma_{1} \sigma_2 -\delta \bar{\delta} } . 
\end{align}
To discuss more  explicitly, we  consider  a   black hole solution with two charges: 
\begin{align}
U(r) &=r^2 -\frac{m_0}{r}-\frac{k ^2}{2} +\frac{1}{4 r^2} \left( \frac{Q_1^2}{Z_0 } +\frac{Q_2^2}{W_0 }  \right) ,
\end{align}
where $m_0$ is given by  $U(r_0)=0$ and  the  temperature   is  
\begin{align}\label{r0T} 
T 
=\frac{ r_0}{4\pi}\left( 3 -\frac{k^2}{2 r_0^{2}}-\frac{ Q_1^2}{4Z_0 r_0^4} -\frac{ Q_2^2}{4 W_0 r_0^4}    \right).
\end{align} 
The solutions of $U(1)$ gauge fields are
 $a(r) =  \mu_1 -\frac{q_1}{r} $, $b(r) = \mu_2  -\frac{q_2}{r}$.
 Notice $q_{i}=Q_{i}/Z_{i}$ with  $Z_{1},Z_{2}$  being $Z_{0}, W_{0}$ respectively.
For the finite vector norm $g^{\mu\nu}A_{\mu}A_{\nu}$  at the horizon $r=r_{0}$,   we need  $\mu_{i}=q_{i}/r_{0}$.  

The conductivities  for any number of conserved  currents   can   be calculated  explicitly:  
\begin{align}
\sigma_{i} &= Z_i +\frac{Q_i^2}{r_0^2 k ^2} , \;\; \sigma_{ij}=\frac{Q_i Q_{j}}{r_0^2 k ^2} 
,\;\;
\kappa = \frac{\bar\kappa}{1+\sum_{i}{4\pi Q_i^2}/sk^2 {Z_i}}, \nonumber
\end{align}
with ${\bar\kappa}=4\pi sT/k^{2}$, $s=4 \pi r_0^2$ and   $Z_i$ is the coupling  of  $A_{i}$.  
 If we identify the total electric current  as $J=\sum_{i} J_{i}$ and 
thermo-electric force  as $E_{i}=E-T\nabla(\mu_{i}/T) $, 
  we can calculate the electric conductivity  to give  
\begin{align}
\sigma=\frac{\partial J}{\partial E}=\sum_{i} \sigma_{i}+\sum_{i,j}\sigma_{ij} =Z + 4\pi Q^{2}/{s k ^2},
\end{align}
where    $Q=\sum_{i} Q_i$ and $Z=\sum_{i} Z_i$, showing the additivity of charge-conjugation-invariant part \cite{Blake:2014yla} of the electric conductivity. If we define   the heat conductivity due to the $i$-th current by $1/\kappa_{i}=1/{\bar\kappa}+Q^{2}_{i}/Z_{i}s^{2}T$,   
then  the heat conductivity formula leads us to {\it additivity of dissipative part of the inverse heat conductivity}. Therefore 
\be 
D[1/\kappa]=\sum_{i}D[1/\kappa_{i}],\quad \quad
{\bar D}[ \sigma]=\sum_{i}{\bar D}[\sigma_{i}],
\label{additivity}\ee 
where  $D[f]$ denotes the dissipative  part of $f$ and ${\bar D}[f]=f-D[f]$ .

Finally we  claim that the experimental data of graphene will be well fit with two current theory  if  we assume the proportionality of   two charges
\begin{align} 
Q_2=g Q_{1} ,\label{qlinearity}
\end{align} 
whose justification will be discussed later. This assumption together with the   additivities in  eq.(\ref{additivity})  are what makes our two current model work.

{\bf    Origin of two Currents in Graphene:}
What is the nature and the origin of the extra  current in the graphene. 
There are a few attractive candidates.
The first idea is the effect of  imbalance \cite{Foster} between the electrons and holes due to the kinematical constraints of the Dirac cone. When there is such an deviation  of electron and hole density from their equilibrium value, then the system has tendency to reduce the difference by 
creating/absorbing electron-hole pair: 
\begin{equation}
e^{-}\leftrightarrow e^{-} + h^{+} +e^{-} , \;\;\;
h^{+}\leftrightarrow h^{+} + h^{+} +e^{-} 
\end{equation}
In such processes, energy and momemtum conservations must hold. The point is that,  
for the  graphene, the linear dispersion relation 
severely reduces the kinematically available states \cite{Foster}: 
If we define $\vec{q}$  as a momentum measured from a Dirac point,  
\begin{align}  
  \vec{q}_{1}=\vec{q}_{2}+\vec{q}_{3}+\vec{q}_{4},\quad |\vec{q_{1}}| = |\vec{q_{2}}|+|\vec{q_{2}}|+|\vec{q_{3}}|,
  \end{align}
    which request the  co-linearity of  all momentum vectors  $\vec{q}_{1},\cdots, \vec{q}_{4}$. 
  Therefore available phase space is greatly reduced.
Such kinematical constraints maintains the non-equilibrium states and as a consequence,  
 the two currents $J_{e}, J_{h}$ behave   independently  for a long time compared with the Planck time $\sim \hbar/kT$, which is the time for hydrodynamics to work. 

The  net electric current  $J$ and total number  current  $J_{n}$ which become neutral at Dirac point,  
are defined by 
$ J =J_{e}+J_{h},    \quad J_{n}=J_{e}-J_{h}, $
respectively and their electric charge densities  and number densities are related by 
$Q_{1}= e n_{1}$ and  $Q_{2}=-e n_{2}$. 
The total electric conductivity $\sigma=\frac{\partial J}{\partial E}$ and $\kappa$ can be 
expressed in terms of 
$Q=Q_{1}+Q_{2} $ and $ Q_{n}=Q_{1}-Q_{2}$ together with the proportionality constant $g_n$ of $Q_{n}=g_{n}Q$:  
\begin{align}
\sigma =\sigma_{0}(1+(Q /Q_{0})^{2}),\;
\kappa=\frac{\bar{\kappa}}{ 1+(1+g_{n}^{2})(Q /Q_{0})^{2}  },
\end{align}
where 
\be
\sigma_{0}= \frac{e^{2}}{\hbar}2Z_{0}, \quad
\bar{\kappa}= \frac{4\pi k_{B}}\hbar    \frac{sT}{k^{2}}, \quad
Q_{0}^{2}=\frac{\hbar\sigma_{0}}{4\pi k_{B}}sk^{2}. 
\ee
To fix the parameters, we used four measured values of ref. \cite{pkim} at 75K, 
 $\sigma_{0}= 0.338/k\Omega$, 
$\bar{\kappa}= 7.7 nW/K$ ,  $Q_{0}=e\cdot 320 /(\mu m)^{2}$, together with 
the curvature of density plot of $\kappa$ to fix $g_{n}=3.2$ 
 and assumed charge conjugation symmetry to set $W_{0}=Z_{0}$.
 Using these, the basic parameters of the theory as well as the entropy density 
 can be determined:
$ 2Z_{0}=1.387, k^{2}={454\over (\mu m)^{2}} , \; s=2044 {k_{B}\over (\mu m)^{2}}
$. We replace all $r_{0}$ dependence by $s$, the entropy density by $s=4\pi k_{B}r_{0}^{2}$. Cosmological constant is not determined due to the inherent scale symmetry.  The resulting fit to data is given in Fig.1.

\begin{figure}
\begin{center}
\subfigure[]{\includegraphics[angle=0,width=0.4\textwidth]{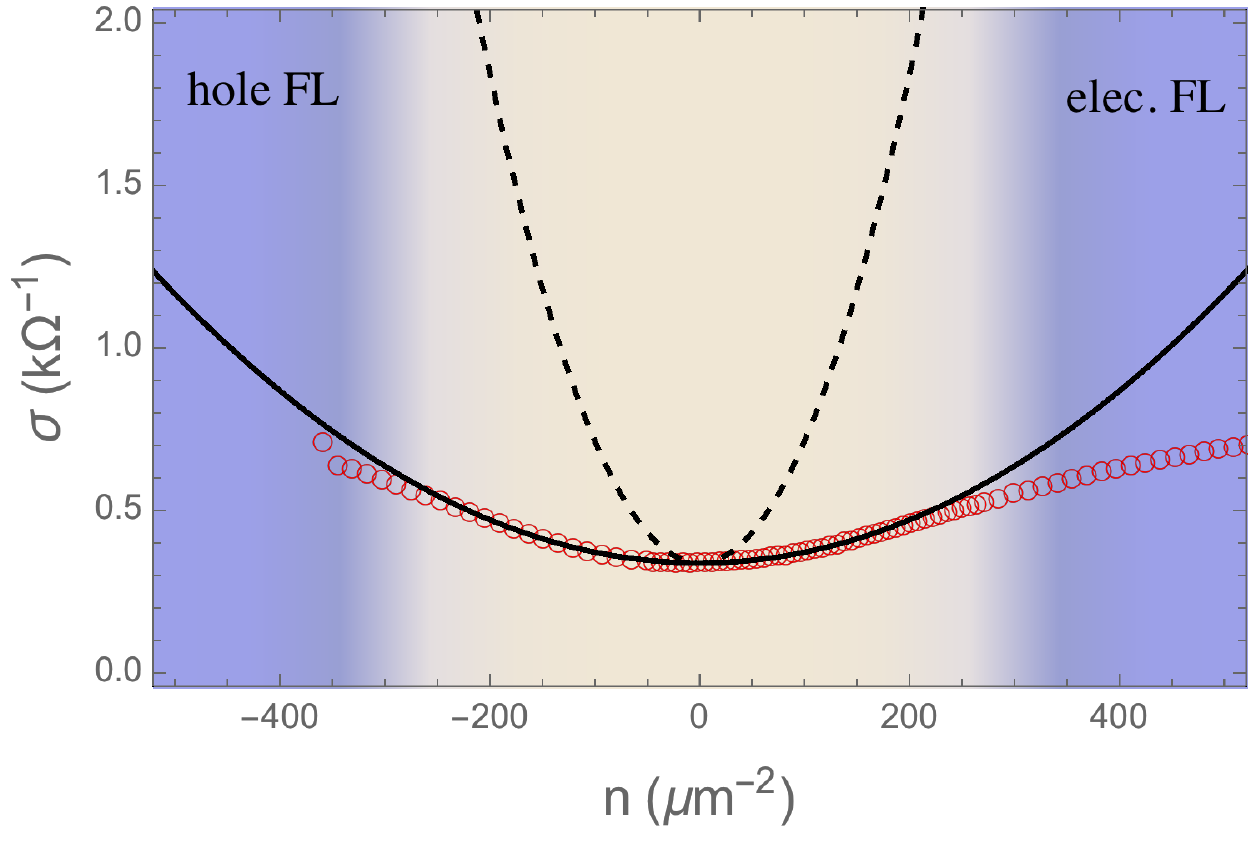}}
\subfigure[]{\includegraphics[angle=0,width=0.4\textwidth]{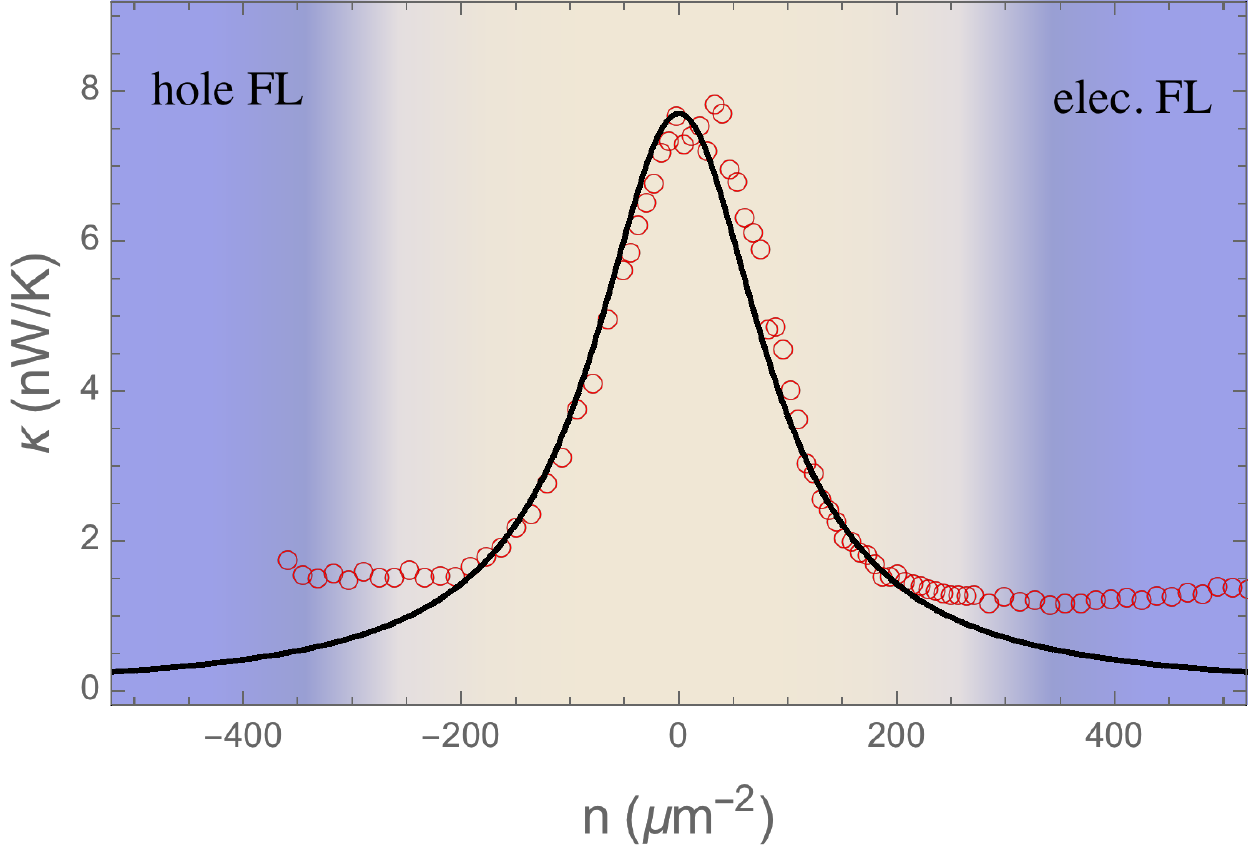}}
\caption{ Theory vs. Data: (a)density plot of  $\sigma$, (b)that of $\kappa$. Red circles are for data used in \cite{pkim,Lucas:2015sya},  dashed lines are for one current model and real lines are for two current model.   The blue  color  is 
for the Fermi Liquid regime and near charge neutral point is the Dirac fluid regime, where our  theory works. Color grading is given to guide the eye. 
   } \label{fig} 
\end{center}
\end{figure}

Now, why we can set the {\it proportionality of the two charges } as given in eq. (\ref{qlinearity}). 
%
%
To avoid the issues involved in the transport by puddle,  we simply  assume that the system is homogeneous. Then the number densities of electrons and holes created by thermal excitation is proportional to the net charge density:
for the fermi liquid case, out of total degree of freedom (d.o.f) $n \sim k_{F}^{2} \sim \mu^{2}$, excitable d.o.f is $\sim  kT \cdot \mu$,  because  the excitable shell width is $kT$. But in hydrodynamic regime, 
 $kT >> \mu$, therefore entire non-degenerate charge distribution region is excitable. In fact this is a typical situation of fermion dynamics described by AdS black hole \cite{sslee,Liu:2011aa}. In summary,
in case of the hydrodynamic regime, the charge carrier density created is proportional to total degree of freedom, Q, which   is  the volume of the Dirac cone above the Dirac point.  

We remark that due to strong Coulomb interaction, the created electron hole pairs can form the bound state, exciton. Such excitons in homogeneous graphene satisfies the linear relations between the electric charge and the exciton number. Although exciton in graphene has been discussed extensively \cite{exciton1, PhysRevB.78.121401}, mosts are only for bi-layer graphene. However, we expect that strong coulomb interaction in Dirac Fluid regime of single layer graphene also should  be able to make bound state.
 
\vskip .2cm
{\bf  Discussions:}
 In the presence of an extra current that carries mainly heat, the violation of WFL is not direct evidence of a Dirac fluid. However, the fact that such a phenomenon is quantitatively well 
described by hydrodynamics and gauge/gravity duality, indicates that the system is strongly correlated. 

  { \it  Disorder and the nature of the scalar field:} 
The scalar field provides momentum dissipation only when both its gradient and the vacuum expectation value of its dual operator, $\left<{{\cal O}_{I}}\right>$, are nonzero. The latter is the analogue of charge density in electric field as one can see from the Ward identity,
   \begin{align}
\nabla_{\nu} T^{\mu\nu}= \left<{{\cal O}_{I}}\right> \nabla_{\mu}\chi_{I}^{0} +F^{0}_{\mu\nu}\left<{J^{\nu}}\right> .
\end{align}
The role of the source field $\chi_{I}^{0}=kx_I$ is the chemical potential of impurity 
and that of $\vev{{\cal O}_{I}}$ is the density  of impurity, whose  presence
gives the momentum dissipation. It is identified as the subleading order term of the fluctuation of the scalar field  near the boundary and nonzero  due to the  presence of curvature in AdS spacetime. 
 $k^{2}$ can be understood as the   density of the uniform  impurity.  
  
{\it Other origins of the second current:} We suggested imbalacnce   and   excitons
as possible sources for the extra  current. Here   we discuss   other candidates. 
 {\it i) Spin charge separation}:This is the simplest to explain the phenomena if such separation could  be experimentally confirmed: the spinons are obviously the chargeless heat carrier and densities of spinnons and  holons must be the same and equal to the original electron density. 
  {\it  ii) valley currents}: Graphene consists of two sublattices A and B and such electrons in each sublattice do not scatter, hence they form two  conserved currents. 
However, they  do not necessarily satisfy the linearity condition eq.(\ref{qlinearity}).
 {\it iii) Phonon}: At high   temperature, the phonons are   the main heat carriers in carbon materials. However,  there are good reasons  that phonon is not the main player in the   regime we are discussing \cite{pkim}. 

 {\it Future directions}:
  It would be  interesting if we can extend our method  to multilayered graphene  and graphite.   Some holographic analysis for the latter  was already reported \cite{kkss}. 
  The thermo-electric power and  magneto-transport are also very interesting observable    for the Dirac Fluid regime. We note that some of the early data    began to be produced\cite{Ghahari:2016ab}. 
  From the experimental side, the abundance of   excitons in single layer graphene is remained to be verified experimentally.  
   The strong correlation was measured only in the limited temperature  window 
$45K<T<90K$ and in the density regime near the charge neutral point\cite{pkim}, outside of which graphene  has well been described as weekly interacting system whose gravity dual is hard to find if exist.  Nevertheless  presence of such strongly interacting regime can give us  extraordinary  guide in constructing the general theory of strongly interacting many body system.

 \begin{acknowledgments}
We would like to thank  Mathew Foster and Youngwoo Son for helpful discussions. 
 This  work is supported by Mid-career Researcher Program through the National Research Foundation of Korea grant No. NRF-2016R1A2B3007687.  
 PK acknowledges the support from the Gordon and Betty Moore Foundation’s EPiQS Initiative through Grant GBMF4543. SJS thanks to apctp for the support through the focus program.  SS was supported by MURI grant W911NF-14-1-0003 from ARO.
Research at Perimeter Institute is supported by the Government of Canada through Industry Canada and by the Province of Ontario 
through the Ministry of Research and Innovation. SS also acknowledges support 
from Cenovus Energy at Perimeter Institute. 
\end{acknowledgments}


\bibliography{Refs}

\begin{thebibliography}{32}%
\makeatletter
\providecommand \@ifxundefined [1]{%
 \@ifx{#1\undefined}
}%
\providecommand \@ifnum [1]{%
 \ifnum #1\expandafter \@firstoftwo
 \else \expandafter \@secondoftwo
 \fi
}%
\providecommand \@ifx [1]{%
 \ifx #1\expandafter \@firstoftwo
 \else \expandafter \@secondoftwo
 \fi
}%
\providecommand \natexlab [1]{#1}%
\providecommand \enquote  [1]{``#1''}%
\providecommand \bibnamefont  [1]{#1}%
\providecommand \bibfnamefont [1]{#1}%
\providecommand \citenamefont [1]{#1}%
\providecommand \href@noop [0]{\@secondoftwo}%
\providecommand \href [0]{\begingroup \@sanitize@url \@href}%
\providecommand \@href[1]{\@@startlink{#1}\@@href}%
\providecommand \@@href[1]{\endgroup#1\@@endlink}%
\providecommand \@sanitize@url [0]{\catcode `\\12\catcode `\$12\catcode
  `\&12\catcode `\#12\catcode `\^12\catcode `\_12\catcode `\%12\relax}%
\providecommand \@@startlink[1]{}%
\providecommand \@@endlink[0]{}%
\providecommand \url  [0]{\begingroup\@sanitize@url \@url }%
\providecommand \@url [1]{\endgroup\@href {#1}{\urlprefix }}%
\providecommand \urlprefix  [0]{URL }%
\providecommand \Eprint [0]{\href }%
\providecommand \doibase [0]{http://dx.doi.org/}%
\providecommand \selectlanguage [0]{\@gobble}%
\providecommand \bibinfo  [0]{\@secondoftwo}%
\providecommand \bibfield  [0]{\@secondoftwo}%
\providecommand \translation [1]{[#1]}%
\providecommand \BibitemOpen [0]{}%
\providecommand \bibitemStop [0]{}%
\providecommand \bibitemNoStop [0]{.\EOS\space}%
\providecommand \EOS [0]{\spacefactor3000\relax}%
\providecommand \BibitemShut  [1]{\csname bibitem#1\endcsname}%
\let\auto@bib@innerbib\@empty
\bibitem [{\citenamefont {{M{\"u}ller}}\ and\ \citenamefont
  {{Sachdev}}(2008)}]{MMSS08}%
  \BibitemOpen
  \bibfield  {author} {\bibinfo {author} {\bibfnamefont {M.}~\bibnamefont
  {{M{\"u}ller}}}\ and\ \bibinfo {author} {\bibfnamefont {S.}~\bibnamefont
  {{Sachdev}}},\ }\href {\doibase 10.1103/PhysRevB.78.115419} {\bibfield
  {journal} {\bibinfo  {journal} {Phys. Rev. B}\ }\textbf {\bibinfo {volume}
  {78}},\ \bibinfo {eid} {115419} (\bibinfo {year} {2008})},\ \Eprint
  {http://arxiv.org/abs/0801.2970} {arXiv:0801.2970 [cond-mat.str-el]}
  \BibitemShut {NoStop}%
\bibitem [{\citenamefont {{Fritz}}\ \emph {et~al.}(2008)\citenamefont
  {{Fritz}}, \citenamefont {{Schmalian}}, \citenamefont {{M{\"u}ller}},\ and\
  \citenamefont {{Sachdev}}}]{FSMS08}%
  \BibitemOpen
  \bibfield  {author} {\bibinfo {author} {\bibfnamefont {L.}~\bibnamefont
  {{Fritz}}}, \bibinfo {author} {\bibfnamefont {J.}~\bibnamefont
  {{Schmalian}}}, \bibinfo {author} {\bibfnamefont {M.}~\bibnamefont
  {{M{\"u}ller}}}, \ and\ \bibinfo {author} {\bibfnamefont {S.}~\bibnamefont
  {{Sachdev}}},\ }\href {\doibase 10.1103/PhysRevB.78.085416} {\bibfield
  {journal} {\bibinfo  {journal} {Phys. Rev. B}\ }\textbf {\bibinfo {volume}
  {78}},\ \bibinfo {eid} {085416} (\bibinfo {year} {2008})},\ \Eprint
  {http://arxiv.org/abs/0802.4289} {arXiv:0802.4289} \BibitemShut {NoStop}%
\bibitem [{\citenamefont {{M{\"u}ller}}\ \emph {et~al.}(2008)\citenamefont
  {{M{\"u}ller}}, \citenamefont {{Fritz}},\ and\ \citenamefont
  {{Sachdev}}}]{MFS08}%
  \BibitemOpen
  \bibfield  {author} {\bibinfo {author} {\bibfnamefont {M.}~\bibnamefont
  {{M{\"u}ller}}}, \bibinfo {author} {\bibfnamefont {L.}~\bibnamefont
  {{Fritz}}}, \ and\ \bibinfo {author} {\bibfnamefont {S.}~\bibnamefont
  {{Sachdev}}},\ }\href {\doibase 10.1103/PhysRevB.78.115406} {\bibfield
  {journal} {\bibinfo  {journal} {Phys. Rev. B}\ }\textbf {\bibinfo {volume}
  {78}},\ \bibinfo {eid} {115406} (\bibinfo {year} {2008})},\ \Eprint
  {http://arxiv.org/abs/0805.1413} {arXiv:0805.1413 [cond-mat.str-el]}
  \BibitemShut {NoStop}%
\bibitem [{\citenamefont {{Foster}}\ and\ \citenamefont
  {{Aleiner}}(2009)}]{Foster}%
  \BibitemOpen
  \bibfield  {author} {\bibinfo {author} {\bibfnamefont {M.~S.}\ \bibnamefont
  {{Foster}}}\ and\ \bibinfo {author} {\bibfnamefont {I.~L.}\ \bibnamefont
  {{Aleiner}}},\ }\href {\doibase 10.1103/PhysRevB.79.085415} {\bibfield
  {journal} {\bibinfo  {journal} {Physical Review B}\ }\textbf {\bibinfo
  {volume} {79}},\ \bibinfo {eid} {085415} (\bibinfo {year} {2009})},\ \Eprint
  {http://arxiv.org/abs/0810.4342} {arXiv:0810.4342 [cond-mat.mes-hall]}
  \BibitemShut {NoStop}%
\bibitem [{\citenamefont {M\"uller}\ \emph {et~al.}(2009)\citenamefont
  {M\"uller}, \citenamefont {Schmalian},\ and\ \citenamefont {Fritz}}]{MSF09}%
  \BibitemOpen
  \bibfield  {author} {\bibinfo {author} {\bibfnamefont {M.}~\bibnamefont
  {M\"uller}}, \bibinfo {author} {\bibfnamefont {J.}~\bibnamefont {Schmalian}},
  \ and\ \bibinfo {author} {\bibfnamefont {L.}~\bibnamefont {Fritz}},\ }\href
  {\doibase 10.1103/PhysRevLett.103.025301} {\bibfield  {journal} {\bibinfo
  {journal} {Phys. Rev. Lett.}\ }\textbf {\bibinfo {volume} {103}},\ \bibinfo
  {pages} {025301} (\bibinfo {year} {2009})},\ \Eprint
  {http://arxiv.org/abs/0903.4178} {arXiv:0903.4178 [cond-mat.mes-hall]}
  \BibitemShut {NoStop}%
\bibitem [{\citenamefont {Mendoza}\ \emph {et~al.}(2011)\citenamefont
  {Mendoza}, \citenamefont {Herrmann},\ and\ \citenamefont
  {Succi}}]{Mendoza11}%
  \BibitemOpen
  \bibfield  {author} {\bibinfo {author} {\bibfnamefont {M.}~\bibnamefont
  {Mendoza}}, \bibinfo {author} {\bibfnamefont {H.~J.}\ \bibnamefont
  {Herrmann}}, \ and\ \bibinfo {author} {\bibfnamefont {S.}~\bibnamefont
  {Succi}},\ }\href {\doibase 10.1103/PhysRevLett.106.156601} {\bibfield
  {journal} {\bibinfo  {journal} {Phys. Rev. Lett.}\ }\textbf {\bibinfo
  {volume} {106}},\ \bibinfo {pages} {156601} (\bibinfo {year} {2011})},\
  \Eprint {http://arxiv.org/abs/1201.6590} {arXiv:1201.6590
  [cond-mat.mes-hall]} \BibitemShut {NoStop}%
\bibitem [{\citenamefont {{Tomadin}}\ \emph {et~al.}(2014)\citenamefont
  {{Tomadin}}, \citenamefont {{Vignale}},\ and\ \citenamefont
  {{Polini}}}]{Polini14}%
  \BibitemOpen
  \bibfield  {author} {\bibinfo {author} {\bibfnamefont {A.}~\bibnamefont
  {{Tomadin}}}, \bibinfo {author} {\bibfnamefont {G.}~\bibnamefont
  {{Vignale}}}, \ and\ \bibinfo {author} {\bibfnamefont {M.}~\bibnamefont
  {{Polini}}},\ }\href {\doibase 10.1103/PhysRevLett.113.235901} {\bibfield
  {journal} {\bibinfo  {journal} {Phys. Rev. Lett.}\ }\textbf {\bibinfo
  {volume} {113}},\ \bibinfo {eid} {235901} (\bibinfo {year} {2014})},\ \Eprint
  {http://arxiv.org/abs/1401.0938} {arXiv:1401.0938 [cond-mat.mes-hall]}
  \BibitemShut {NoStop}%
\bibitem [{\citenamefont {{Principi}}\ and\ \citenamefont
  {{Vignale}}(2015)}]{Vignale15}%
  \BibitemOpen
  \bibfield  {author} {\bibinfo {author} {\bibfnamefont {A.}~\bibnamefont
  {{Principi}}}\ and\ \bibinfo {author} {\bibfnamefont {G.}~\bibnamefont
  {{Vignale}}},\ }\href {\doibase 10.1103/PhysRevLett.115.056603} {\bibfield
  {journal} {\bibinfo  {journal} {Phys. Rev. Lett.}\ }\textbf {\bibinfo
  {volume} {115}},\ \bibinfo {eid} {056603} (\bibinfo {year} {2015})},\ \Eprint
  {http://arxiv.org/abs/1406.2940} {arXiv:1406.2940 [cond-mat.mes-hall]}
  \BibitemShut {NoStop}%
\bibitem [{\citenamefont {{Torre}}\ \emph {et~al.}(2015)\citenamefont
  {{Torre}}, \citenamefont {{Tomadin}}, \citenamefont {{Geim}},\ and\
  \citenamefont {{Polini}}}]{Polini15}%
  \BibitemOpen
  \bibfield  {author} {\bibinfo {author} {\bibfnamefont {I.}~\bibnamefont
  {{Torre}}}, \bibinfo {author} {\bibfnamefont {A.}~\bibnamefont {{Tomadin}}},
  \bibinfo {author} {\bibfnamefont {A.~K.}\ \bibnamefont {{Geim}}}, \ and\
  \bibinfo {author} {\bibfnamefont {M.}~\bibnamefont {{Polini}}},\ }\href
  {\doibase 10.1103/PhysRevB.92.165433} {\bibfield  {journal} {\bibinfo
  {journal} {Phys. Rev. B}\ }\textbf {\bibinfo {volume} {92}},\ \bibinfo {eid}
  {165433} (\bibinfo {year} {2015})},\ \Eprint
  {http://arxiv.org/abs/1508.00363} {arXiv:1508.00363 [cond-mat.mes-hall]}
  \BibitemShut {NoStop}%
\bibitem [{\citenamefont {Levitov}\ and\ \citenamefont
  {Falkovich}(2016)}]{Levitov16}%
  \BibitemOpen
  \bibfield  {author} {\bibinfo {author} {\bibfnamefont {L.}~\bibnamefont
  {Levitov}}\ and\ \bibinfo {author} {\bibfnamefont {G.}~\bibnamefont
  {Falkovich}},\ }\href {http://dx.doi.org/10.1038/nphys3667} {\bibfield
  {journal} {\bibinfo  {journal} {Nat Phys}\ }\textbf {\bibinfo {volume}
  {12}},\ \bibinfo {pages} {672} (\bibinfo {year} {2016})},\ \Eprint
  {http://arxiv.org/abs/1508.00836} {arXiv:1508.00836 [cond-mat.mes-hall]}
  \BibitemShut {NoStop}%
\bibitem [{\citenamefont {{Crossno}}\ \emph {et~al.}(2016)\citenamefont
  {{Crossno}}, \citenamefont {{Shi}}, \citenamefont {{Wang}}, \citenamefont
  {{Liu}}, \citenamefont {{Harzheim}}, \citenamefont {{Lucas}}, \citenamefont
  {{Sachdev}}, \citenamefont {{Kim}}, \citenamefont {{Taniguchi}},
  \citenamefont {{Watanabe}}, \citenamefont {{Ohki}},\ and\ \citenamefont
  {{Fong}}}]{pkim}%
  \BibitemOpen
  \bibfield  {author} {\bibinfo {author} {\bibfnamefont {J.}~\bibnamefont
  {{Crossno}}}, \bibinfo {author} {\bibfnamefont {J.~K.}\ \bibnamefont
  {{Shi}}}, \bibinfo {author} {\bibfnamefont {K.}~\bibnamefont {{Wang}}},
  \bibinfo {author} {\bibfnamefont {X.}~\bibnamefont {{Liu}}}, \bibinfo
  {author} {\bibfnamefont {A.}~\bibnamefont {{Harzheim}}}, \bibinfo {author}
  {\bibfnamefont {A.}~\bibnamefont {{Lucas}}}, \bibinfo {author} {\bibfnamefont
  {S.}~\bibnamefont {{Sachdev}}}, \bibinfo {author} {\bibfnamefont
  {P.}~\bibnamefont {{Kim}}}, \bibinfo {author} {\bibfnamefont
  {T.}~\bibnamefont {{Taniguchi}}}, \bibinfo {author} {\bibfnamefont
  {K.}~\bibnamefont {{Watanabe}}}, \bibinfo {author} {\bibfnamefont {T.~A.}\
  \bibnamefont {{Ohki}}}, \ and\ \bibinfo {author} {\bibfnamefont {K.~C.}\
  \bibnamefont {{Fong}}},\ }\href {\doibase 10.1126/science.aad0343} {\bibfield
   {journal} {\bibinfo  {journal} {Science}\ }\textbf {\bibinfo {volume}
  {351}},\ \bibinfo {pages} {1058} (\bibinfo {year} {2016})},\ \Eprint
  {http://arxiv.org/abs/1509.04713} {arXiv:1509.04713 [cond-mat.mes-hall]}
  \BibitemShut {NoStop}%
\bibitem [{\citenamefont {{Hartnoll}}\ \emph {et~al.}(2007)\citenamefont
  {{Hartnoll}}, \citenamefont {{Kovtun}}, \citenamefont {{M{\"u}ller}},\ and\
  \citenamefont {{Sachdev}}}]{HKMS}%
  \BibitemOpen
  \bibfield  {author} {\bibinfo {author} {\bibfnamefont {S.~A.}\ \bibnamefont
  {{Hartnoll}}}, \bibinfo {author} {\bibfnamefont {P.~K.}\ \bibnamefont
  {{Kovtun}}}, \bibinfo {author} {\bibfnamefont {M.}~\bibnamefont
  {{M{\"u}ller}}}, \ and\ \bibinfo {author} {\bibfnamefont {S.}~\bibnamefont
  {{Sachdev}}},\ }\href {\doibase 10.1103/PhysRevB.76.144502} {\bibfield
  {journal} {\bibinfo  {journal} {Phys. Rev. B}\ }\textbf {\bibinfo {volume}
  {76}},\ \bibinfo {eid} {144502} (\bibinfo {year} {2007})},\ \Eprint
  {http://arxiv.org/abs/0706.3215} {arXiv:0706.3215 [cond-mat.str-el]}
  \BibitemShut {NoStop}%
\bibitem [{\citenamefont {Lucas}\ \emph {et~al.}(2016)\citenamefont {Lucas},
  \citenamefont {Crossno}, \citenamefont {Fong}, \citenamefont {Kim},\ and\
  \citenamefont {Sachdev}}]{Lucas:2015sya}%
  \BibitemOpen
  \bibfield  {author} {\bibinfo {author} {\bibfnamefont {A.}~\bibnamefont
  {Lucas}}, \bibinfo {author} {\bibfnamefont {J.}~\bibnamefont {Crossno}},
  \bibinfo {author} {\bibfnamefont {K.~C.}\ \bibnamefont {Fong}}, \bibinfo
  {author} {\bibfnamefont {P.}~\bibnamefont {Kim}}, \ and\ \bibinfo {author}
  {\bibfnamefont {S.}~\bibnamefont {Sachdev}},\ }\href {\doibase
  10.1103/PhysRevB.93.075426} {\bibfield  {journal} {\bibinfo  {journal} {Phys.
  Rev.}\ }\textbf {\bibinfo {volume} {B93}},\ \bibinfo {pages} {075426}
  (\bibinfo {year} {2016})},\ \Eprint {http://arxiv.org/abs/1510.01738}
  {arXiv:1510.01738 [cond-mat.str-el]} \BibitemShut {NoStop}%
\bibitem [{\citenamefont {Maldacena}(1999)}]{Maldacena:1997re}%
  \BibitemOpen
  \bibfield  {author} {\bibinfo {author} {\bibfnamefont {J.~M.}\ \bibnamefont
  {Maldacena}},\ }\href {\doibase 10.1023/A:1026654312961} {\bibfield
  {journal} {\bibinfo  {journal} {Int.J.Theor.Phys.}\ }\textbf {\bibinfo
  {volume} {38}},\ \bibinfo {pages} {1113} (\bibinfo {year} {1999})},\ \Eprint
  {http://arxiv.org/abs/hep-th/9711200} {arXiv:hep-th/9711200 [hep-th]}
  \BibitemShut {NoStop}%
\bibitem [{\citenamefont {Witten}(1998)}]{Witten:1998qj}%
  \BibitemOpen
  \bibfield  {author} {\bibinfo {author} {\bibfnamefont {E.}~\bibnamefont
  {Witten}},\ }\href@noop {} {\bibfield  {journal} {\bibinfo  {journal} {Adv.
  Theor. Math. Phys.}\ }\textbf {\bibinfo {volume} {2}},\ \bibinfo {pages}
  {253} (\bibinfo {year} {1998})},\ \Eprint
  {http://arxiv.org/abs/hep-th/9802150} {arXiv:hep-th/9802150} \BibitemShut
  {NoStop}%
\bibitem [{\citenamefont {Blake}\ \emph {et~al.}(2014)\citenamefont {Blake},
  \citenamefont {Tong},\ and\ \citenamefont {Vegh}}]{Blake:2013owa}%
  \BibitemOpen
  \bibfield  {author} {\bibinfo {author} {\bibfnamefont {M.}~\bibnamefont
  {Blake}}, \bibinfo {author} {\bibfnamefont {D.}~\bibnamefont {Tong}}, \ and\
  \bibinfo {author} {\bibfnamefont {D.}~\bibnamefont {Vegh}},\ }\href {\doibase
  10.1103/PhysRevLett.112.071602} {\bibfield  {journal} {\bibinfo  {journal}
  {Phys. Rev. Lett.}\ }\textbf {\bibinfo {volume} {112}},\ \bibinfo {pages}
  {071602} (\bibinfo {year} {2014})},\ \Eprint {http://arxiv.org/abs/1310.3832}
  {arXiv:1310.3832 [hep-th]} \BibitemShut {NoStop}%
\bibitem [{\citenamefont {Andrade}\ and\ \citenamefont
  {Withers}(2014)}]{Andrade:2013gsa}%
  \BibitemOpen
  \bibfield  {author} {\bibinfo {author} {\bibfnamefont {T.}~\bibnamefont
  {Andrade}}\ and\ \bibinfo {author} {\bibfnamefont {B.}~\bibnamefont
  {Withers}},\ }\href {\doibase 10.1007/JHEP05(2014)101} {\bibfield  {journal}
  {\bibinfo  {journal} {JHEP}\ }\textbf {\bibinfo {volume} {1405}},\ \bibinfo
  {pages} {101} (\bibinfo {year} {2014})},\ \Eprint
  {http://arxiv.org/abs/1311.5157} {arXiv:1311.5157 [hep-th]} \BibitemShut
  {NoStop}%
\bibitem [{\citenamefont {Donos}\ and\ \citenamefont
  {Gauntlett}(2014)}]{Donos:2014cya}%
  \BibitemOpen
  \bibfield  {author} {\bibinfo {author} {\bibfnamefont {A.}~\bibnamefont
  {Donos}}\ and\ \bibinfo {author} {\bibfnamefont {J.~P.}\ \bibnamefont
  {Gauntlett}},\ }\href {\doibase 10.1007/JHEP11(2014)081} {\bibfield
  {journal} {\bibinfo  {journal} {JHEP}\ }\textbf {\bibinfo {volume} {1411}},\
  \bibinfo {pages} {081} (\bibinfo {year} {2014})},\ \Eprint
  {http://arxiv.org/abs/1406.4742} {arXiv:1406.4742 [hep-th]} \BibitemShut
  {NoStop}%
\bibitem [{\citenamefont {Donos}\ and\ \citenamefont
  {Gauntlett}(2015)}]{Donos:2014yya}%
  \BibitemOpen
  \bibfield  {author} {\bibinfo {author} {\bibfnamefont {A.}~\bibnamefont
  {Donos}}\ and\ \bibinfo {author} {\bibfnamefont {J.~P.}\ \bibnamefont
  {Gauntlett}},\ }\href {\doibase 10.1007/JHEP01(2015)035} {\bibfield
  {journal} {\bibinfo  {journal} {JHEP}\ }\textbf {\bibinfo {volume} {01}},\
  \bibinfo {pages} {035} (\bibinfo {year} {2015})},\ \Eprint
  {http://arxiv.org/abs/1409.6875} {arXiv:1409.6875 [hep-th]} \BibitemShut
  {NoStop}%
\bibitem [{\citenamefont {Donos}\ \emph {et~al.}(2014)\citenamefont {Donos},
  \citenamefont {Goutéraux},\ and\ \citenamefont {Kiritsis}}]{Donos:2014oha}%
  \BibitemOpen
  \bibfield  {author} {\bibinfo {author} {\bibfnamefont {A.}~\bibnamefont
  {Donos}}, \bibinfo {author} {\bibfnamefont {B.}~\bibnamefont {Goutéraux}}, \
  and\ \bibinfo {author} {\bibfnamefont {E.}~\bibnamefont {Kiritsis}},\ }\href
  {\doibase 10.1007/JHEP09(2014)038} {\bibfield  {journal} {\bibinfo  {journal}
  {JHEP}\ }\textbf {\bibinfo {volume} {09}},\ \bibinfo {pages} {038} (\bibinfo
  {year} {2014})},\ \Eprint {http://arxiv.org/abs/1406.6351} {arXiv:1406.6351
  [hep-th]} \BibitemShut {NoStop}%
\bibitem [{\citenamefont {Kim}\ \emph {et~al.}(2014)\citenamefont {Kim},
  \citenamefont {Kim}, \citenamefont {Seo},\ and\ \citenamefont
  {Sin}}]{Kim:2014bza}%
  \BibitemOpen
  \bibfield  {author} {\bibinfo {author} {\bibfnamefont {K.-Y.}\ \bibnamefont
  {Kim}}, \bibinfo {author} {\bibfnamefont {K.~K.}\ \bibnamefont {Kim}},
  \bibinfo {author} {\bibfnamefont {Y.}~\bibnamefont {Seo}}, \ and\ \bibinfo
  {author} {\bibfnamefont {S.-J.}\ \bibnamefont {Sin}},\ }\href {\doibase
  10.1007/JHEP12(2014)170} {\bibfield  {journal} {\bibinfo  {journal} {JHEP}\
  }\textbf {\bibinfo {volume} {1412}},\ \bibinfo {pages} {170} (\bibinfo {year}
  {2014})},\ \Eprint {http://arxiv.org/abs/1409.8346} {arXiv:1409.8346
  [hep-th]} \BibitemShut {NoStop}%
\bibitem [{\citenamefont {Ge}\ \emph {et~al.}(2015)\citenamefont {Ge},
  \citenamefont {Ling}, \citenamefont {Niu},\ and\ \citenamefont
  {Sin}}]{Ge:2014aza}%
  \BibitemOpen
  \bibfield  {author} {\bibinfo {author} {\bibfnamefont {X.-H.}\ \bibnamefont
  {Ge}}, \bibinfo {author} {\bibfnamefont {Y.}~\bibnamefont {Ling}}, \bibinfo
  {author} {\bibfnamefont {C.}~\bibnamefont {Niu}}, \ and\ \bibinfo {author}
  {\bibfnamefont {S.-J.}\ \bibnamefont {Sin}},\ }\href {\doibase
  10.1103/PhysRevD.92.106005} {\bibfield  {journal} {\bibinfo  {journal} {Phys.
  Rev.}\ }\textbf {\bibinfo {volume} {D92}},\ \bibinfo {pages} {106005}
  (\bibinfo {year} {2015})},\ \Eprint {http://arxiv.org/abs/1412.8346}
  {arXiv:1412.8346 [hep-th]} \BibitemShut {NoStop}%
\bibitem [{\citenamefont {Kim}\ \emph {et~al.}(2015{\natexlab{a}})\citenamefont
  {Kim}, \citenamefont {Kim}, \citenamefont {Seo},\ and\ \citenamefont
  {Sin}}]{Kim:2015sma}%
  \BibitemOpen
  \bibfield  {author} {\bibinfo {author} {\bibfnamefont {K.-Y.}\ \bibnamefont
  {Kim}}, \bibinfo {author} {\bibfnamefont {K.~K.}\ \bibnamefont {Kim}},
  \bibinfo {author} {\bibfnamefont {Y.}~\bibnamefont {Seo}}, \ and\ \bibinfo
  {author} {\bibfnamefont {S.-J.}\ \bibnamefont {Sin}},\ }\href {\doibase
  10.1016/j.physletb.2015.07.058} {\bibfield  {journal} {\bibinfo  {journal}
  {Phys. Lett.}\ }\textbf {\bibinfo {volume} {B749}},\ \bibinfo {pages} {108}
  (\bibinfo {year} {2015}{\natexlab{a}})},\ \Eprint
  {http://arxiv.org/abs/1502.02100} {arXiv:1502.02100 [hep-th]} \BibitemShut
  {NoStop}%
\bibitem [{\citenamefont {Kim}\ \emph {et~al.}(2015{\natexlab{b}})\citenamefont
  {Kim}, \citenamefont {Kim}, \citenamefont {Seo},\ and\ \citenamefont
  {Sin}}]{Kim:2015wba}%
  \BibitemOpen
  \bibfield  {author} {\bibinfo {author} {\bibfnamefont {K.-Y.}\ \bibnamefont
  {Kim}}, \bibinfo {author} {\bibfnamefont {K.~K.}\ \bibnamefont {Kim}},
  \bibinfo {author} {\bibfnamefont {Y.}~\bibnamefont {Seo}}, \ and\ \bibinfo
  {author} {\bibfnamefont {S.-J.}\ \bibnamefont {Sin}},\ }\href {\doibase
  10.1007/JHEP07(2015)027} {\bibfield  {journal} {\bibinfo  {journal} {JHEP}\
  }\textbf {\bibinfo {volume} {07}},\ \bibinfo {pages} {027} (\bibinfo {year}
  {2015}{\natexlab{b}})},\ \Eprint {http://arxiv.org/abs/1502.05386}
  {arXiv:1502.05386 [hep-th]} \BibitemShut {NoStop}%
\bibitem [{\citenamefont {Blake}\ \emph {et~al.}(2015)\citenamefont {Blake},
  \citenamefont {Donos},\ and\ \citenamefont {Lohitsiri}}]{Blake:2015ina}%
  \BibitemOpen
  \bibfield  {author} {\bibinfo {author} {\bibfnamefont {M.}~\bibnamefont
  {Blake}}, \bibinfo {author} {\bibfnamefont {A.}~\bibnamefont {Donos}}, \ and\
  \bibinfo {author} {\bibfnamefont {N.}~\bibnamefont {Lohitsiri}},\ }\href
  {\doibase 10.1007/JHEP08(2015)124} {\bibfield  {journal} {\bibinfo  {journal}
  {JHEP}\ }\textbf {\bibinfo {volume} {08}},\ \bibinfo {pages} {124} (\bibinfo
  {year} {2015})},\ \Eprint {http://arxiv.org/abs/1502.03789} {arXiv:1502.03789
  [hep-th]} \BibitemShut {NoStop}%
\bibitem [{\citenamefont {Blake}\ and\ \citenamefont
  {Donos}(2015)}]{Blake:2014yla}%
  \BibitemOpen
  \bibfield  {author} {\bibinfo {author} {\bibfnamefont {M.}~\bibnamefont
  {Blake}}\ and\ \bibinfo {author} {\bibfnamefont {A.}~\bibnamefont {Donos}},\
  }\href {\doibase 10.1103/PhysRevLett.114.021601} {\bibfield  {journal}
  {\bibinfo  {journal} {Phys. Rev. Lett.}\ }\textbf {\bibinfo {volume} {114}},\
  \bibinfo {pages} {021601} (\bibinfo {year} {2015})},\ \Eprint
  {http://arxiv.org/abs/1406.1659} {arXiv:1406.1659 [hep-th]} \BibitemShut
  {NoStop}%
\bibitem [{\citenamefont {Seo}\ \emph {et~al.}(2016)\citenamefont {Seo},
  \citenamefont {Kim}, \citenamefont {Kim},\ and\ \citenamefont {Sin}}]{kkss}%
  \BibitemOpen
  \bibfield  {author} {\bibinfo {author} {\bibfnamefont {Y.}~\bibnamefont
  {Seo}}, \bibinfo {author} {\bibfnamefont {K.-Y.}\ \bibnamefont {Kim}},
  \bibinfo {author} {\bibfnamefont {K.~K.}\ \bibnamefont {Kim}}, \ and\
  \bibinfo {author} {\bibfnamefont {S.-J.}\ \bibnamefont {Sin}},\ }\href
  {\doibase 10.1016/j.physletb.2016.05.059} {\bibfield  {journal} {\bibinfo
  {journal} {Phys. Lett.}\ }\textbf {\bibinfo {volume} {B759}},\ \bibinfo
  {pages} {104} (\bibinfo {year} {2016})},\ \Eprint
  {http://arxiv.org/abs/1512.08916} {arXiv:1512.08916 [hep-th]} \BibitemShut
  {NoStop}%
\bibitem [{\citenamefont {Lee}(2009)}]{sslee}%
  \BibitemOpen
  \bibfield  {author} {\bibinfo {author} {\bibfnamefont {S.-S.}\ \bibnamefont
  {Lee}},\ }\href {\doibase 10.1103/PhysRevD.79.086006} {\bibfield  {journal}
  {\bibinfo  {journal} {Phys. Rev.}\ }\textbf {\bibinfo {volume} {D79}},\
  \bibinfo {pages} {086006} (\bibinfo {year} {2009})},\ \Eprint
  {http://arxiv.org/abs/0809.3402} {arXiv:0809.3402 [hep-th]} \BibitemShut
  {NoStop}%
\bibitem [{\citenamefont {Liu}(2011)}]{Liu:2011aa}%
  \BibitemOpen
  \bibfield  {author} {\bibinfo {author} {\bibfnamefont {H.}~\bibnamefont
  {Liu}},\ }\href {\doibase 10.1103/PhysRevD.83.065029} {\bibfield  {journal}
  {\bibinfo  {journal} {Physical Review D}\ }\textbf {\bibinfo {volume} {83}}
  (\bibinfo {year} {2011}),\ 10.1103/PhysRevD.83.065029}\BibitemShut {NoStop}%
\bibitem [{\citenamefont {Eisenstein}\ and\ \citenamefont
  {MacDonald}(2004)}]{exciton1}%
  \BibitemOpen
  \bibfield  {author} {\bibinfo {author} {\bibfnamefont {J.~P.}\ \bibnamefont
  {Eisenstein}}\ and\ \bibinfo {author} {\bibfnamefont {A.~H.}\ \bibnamefont
  {MacDonald}},\ }\href@noop {} {\bibfield  {journal} {\bibinfo  {journal}
  {Nature}\ }\textbf {\bibinfo {volume} {432}},\ \bibinfo {pages} {691}
  (\bibinfo {year} {2004})}\BibitemShut {NoStop}%
\bibitem [{\citenamefont {Min}\ \emph {et~al.}(2008)\citenamefont {Min},
  \citenamefont {Bistritzer}, \citenamefont {Su},\ and\ \citenamefont
  {MacDonald}}]{PhysRevB.78.121401}%
  \BibitemOpen
  \bibfield  {author} {\bibinfo {author} {\bibfnamefont {H.}~\bibnamefont
  {Min}}, \bibinfo {author} {\bibfnamefont {R.}~\bibnamefont {Bistritzer}},
  \bibinfo {author} {\bibfnamefont {J.-J.}\ \bibnamefont {Su}}, \ and\ \bibinfo
  {author} {\bibfnamefont {A.~H.}\ \bibnamefont {MacDonald}},\ }\href {\doibase
  10.1103/PhysRevB.78.121401} {\bibfield  {journal} {\bibinfo  {journal} {Phys.
  Rev. B}\ }\textbf {\bibinfo {volume} {78}},\ \bibinfo {pages} {121401}
  (\bibinfo {year} {2008})}\BibitemShut {NoStop}%
\bibitem [{\citenamefont {Ghahari}(2016)}]{Ghahari:2016ab}%
  \BibitemOpen
  \bibfield  {author} {\bibinfo {author} {\bibfnamefont {F.}~\bibnamefont
  {Ghahari}},\ }\href {\doibase 10.1103/PhysRevLett.116.136802} {\bibfield
  {journal} {\bibinfo  {journal} {Physical Review Letters}\ }\textbf {\bibinfo
  {volume} {116}} (\bibinfo {year} {2016}),\
  10.1103/PhysRevLett.116.136802}\BibitemShut {NoStop}%
\end{thebibliography}%

\end{document}